\documentclass[showpacs,epsf,twocolumn]{revtex4-1}
\usepackage[bottom]{footmisc}
\usepackage{float}
\usepackage{caption}
\usepackage{color,graphicx}
\usepackage{hyperref}
\begin{document}
\title{Fascinating interplay between Charge Density Wave Order and magnetic field in Non-magnetic Rare-Earth Tritelluride LaTe$_{3}$}
\author{Arnab Pariari$^{1,2}$, Sudipta Koley$^{3}$, Shubhankar Roy$^{1,2}$, Ratnadwip Singha$^{1,2}$, Mukul S. Laad$^{2,4}$, A.  Taraphder$^{5,6}$, Prabhat Mandal$^{1,2}$}
\affiliation{$^1$Saha Institute of Nuclear Physics, 1/AF Bidhannagar, Kolkata 700 064, India}
\affiliation{$^2$Homi Bhabha National Institute, Trombay, Mumbai 400085, India}
\affiliation{$^3$Department of Physics, North Eastern Hill University, Shillong, Meghalaya 793022, India}
\affiliation{$^4$Institute of Mathematical Sciences, Taramani, Chennai 600113, India}
\affiliation{$^5$Department of Physics and Centre for Theoretical Studies, Indian Institute of Technology, Kharagpur 721302, India}
\affiliation{$^6$School of Basic Sciences, Indian Institute of Technology, Mandi 751005, India}
\begin{abstract}
Charge density wave (CDW) states in solids bear an intimate connection to underlying fermiology. Modification of the latter by a suitable perturbation provides an attractive handle to unearth novel CDW states.  Here, we combine extensive magnetotransport experiments and first-principles electronic structure calculations on a non-magnetic tritelluride LaTe$_{3}$ single crystal to uncover phenomena rare in CDW systems: $(i)$ hump-like feature in the temperature dependence of resistivity at low temperature under application of magnetic field, which moves to higher temperature with increasing field strength, $(ii)$ highly anisotropic large transverse magnetoresistance (MR) upon rotation of magnetic field about current parallel to crystallographic c-axis, (iii) anomalously large positive MR with spike-like peaks at characteristic angles when the angle between current and field is varied in the bc-plane, (iv) extreme sensitivity of the angular variation of MR on field and temperature. Moreover, our Hall measurement reveals remarkably high carrier mobility $\sim$ 33000 cm$^{2}$/Vs, which is comparable to that observed in some topological semimetals. These novel observations find a comprehensive explication in our density functional theory (DFT) and dynamical mean field theory (DMFT) calculations that capture field-induced electronic structure modification in LaTe$_{3}$. The band structure theory together with transport calculations suggest the possibility of a second field-induced CDW transition from the field-reconstructed Fermi surface, which qualitatively explains the hump in temperature dependence of resistivity at low temperature. Thus, our study exposes the novel manifestations of the interplay between CDW order and field-induced electronic structure modifications in LaTe$_{3}$, and establishes a new route to tune CDW states by perturbations like magnetic field.
\end{abstract}
\maketitle

\section{Introduction}
Formation of charge density wave (CDW) order in solids spontaneously breaks the discrete translational symmetry of the lattice.  As a direct consequence, the ``normal state'' Fermi surface (FS) is destabilized due to CDW gap opening: in the simplest picture, the Fermi surface is completely obliterated by the opening of a full gap, while a partial gapping of the Fermi surface  characterizes unconventional CDW order (notable examples are transition-metal dichalcogenides (TMDs) \cite{1}).  A direct consequence of partial gapping-out of the FS  in a multi-band system is reduction in strong scattering, leading to enhanced Landau-quasiparticle-like coherence in the CDW state.  This implies lower resistivity and high carrier mobility.

Quasi-two-dimensional rare-earth (R) tritellurides RTe$_{3}$, where R=Y, La-Sm, Gd-Tm, have attracted significant interest due to the existence of CDW order \cite{2,3,4,5,6,7,8,9,10,11,12,13,14,15,16,17,18} in a metal without the complications arising from a proximity to correlation-driven Mott transition and strong attendant magnetic fluctuations. These materials crystallize into a weakly orthorhombic (pseudo-tetragonal) structure consisting of double layers of square-planar Te sheets separated by corrugated RTe slabs \cite{3,12,13,14,16}. The long \textbf{b}-axis is perpendicular to the Te planes. In addition, the two neighboring Te square-net sheets exhibit van der Waals (vdW) gap between them, which allows exfoliation of bulk crystals into thin flakes. It has been established that this family of materials is a fertile ground for intriguing physics by tuning the CDW state through perturbations such as chemical or external pressure \cite{10,11,14,16,17}. Ru \emph{et al.} reported a systematic variation of CDW transition temperature in RTe$_{3}$ (where R = Sm, Gd, Tb, Dy, Ho, Er, and Tm) with increasing lattice parameter \cite{10,11}. A systematic study of CDW stability as a function of lattice
parameter was also reported \cite{4,19}. The CDW state in these materials rapidly suppresses and superconductivity appears with the increase of applied pressure \cite{14,17}. Also, a second CDW transition appears at a much lower temperature for the heavier members of the series, R = Dy, Ho, Er, and Tm \cite{10,11,16,17}. However, no such phenomenon has been reported for LaTe$_{3}$ and other non-magnetic members. LaTe$_{3}$ is the lightest member of this family of compounds, where the CDW transition temperature ($T_{cdw}$) is presumed to be higher than 400 K \cite{13}. LaTe$_{3}$ has been characterized by transmission electron microscopy \cite{4}, NMR \cite{9},  angle-resolved photoemission spectroscopy (ARPES) \cite{12}, band-structure calculation, de Haas-van Alphen oscillations \cite{13}, preliminary electronic transport \cite{8} and thermodynamic property measurements \cite{8}. Hitherto, these studies have sought mainly to understand the CDW-induced lattice modulation and accompanying Fermi surface reconstruction. \\

In the absence of CDW order, the unmodulated  Fermi surface comprises bands formed from the p$_{x}$ and p$_{z}$ orbitals of Te atoms in the square-planar layers: two diamond shaped sheets made up of warped inner and outer layers in the \textbf{ac}-plane due to bilayer splitting \cite{13}. The dispersion is identical along \textbf{a}- and \textbf{c}-axis but minimal along \textbf{b}-axis. The CDW instability opens up large gaps on the parts of Fermi surface showing most favorable propensity for nesting \cite{12}.  Notwithstanding much work recently, tuning of CDW order via modification of fermiology under appropriate perturbations remains a largely unadressed issue. Such an endeavour can help to illuminate the effect of a CDW-reconstructed Fermi surface on the transport properties, and ultimately to help constrain theoretical models for the CDW state itself.  Furthermore, the possibility of modifying or unearthing novel quantum orders by inducing changes in the underlying electronic structure with an appropriate perturbation is an issue of considerable and broad general interest in quantum matter. LaTe$_{3}$ presents the opportunity to avoid the complications due to magnetism, and address these issues in detail in a non-magnetic candidate material, through extensive magnetotransport experiments and density-functional theory (DFT) plus dynamical mean-field theory (DMFT) calculations.\\

\section{Sample preparation and technical details}
Single crystals of LaTe$_{3}$ were grown via tellurium flux technique, similar to previous report \cite{8}. At first, high purity La (Alfa Aesar 99.9\%) and Te (Alfa Aesar 99.99\%) were taken in an alumina crucible with a molar ratio La$_{0.025}$Te$_{0.975}$. Next, the crucible was sealed in a quartz tube under vacuum (10$^{-5}$ Torr). The tube was put in a box furnace vertically and heated to 900$^\circ$C at 60$^\circ$C/h. It was then kept for 12 h and slowly cooled to end temperature 600$^\circ$C over a period of 4 days. Finally, we separated the excess tellurium with a high-temperature centrifuge. Typical length and width of the gold-colored crystals are 4 to 6 mm, and thickness is 0.2 to 0.4 mm. The crystals were freshly cleaved before characterizations and measurements. Phase purity and the structural analysis of the samples were done using high-resolution x-ray diffraction (XRD) in Rigaku, TTRAX II, using Cu-K$_{\alpha}$ radiation. The results of phase purity analysis and structural characterization have been discussed in Section II and Section III of \textbf{Supplementary Materials} and shown in \textbf{Figs. S1 and S2}. Formation of CDW state was confirmed from low-energy electron diffraction study and resistivity measurements across the transition. This has been discussed in detail in \textbf{Supplementary Materials} and shown in \textbf{Figs. S3 and S4}. The resistivity measurements of LaTe$_{3}$ single crystals were done using the standard four-probe technique. Electrical contacts were made using conductive silver paste and thin gold wire. The transport measurements were carried out in a 9-T physical property measurement system (Quantum Design). The magnetization was measured using a 7-T MPMS3 (Quantum Design). Although, several single crystals have been studied, we present the data for a representative single crystal. Qualitative similar behavior has been observed for other crystals.\\

We combine density functional theory  and dynamical mean field theory  in order to examine the electronic structure and resulting properties of LaTe$_{3}$. Calculations are based on the experimentally determined space group symmetry (Cmcm) and lattice parameters, as described in the present and earlier studies \cite{12,13}. DFT calculations for LaTe$_{3}$ have been performed using the WIEN2k full-potential linearized augmented plane wave (FP-LAPW) ab initio package \cite{20,21}.  Using Multi-Orbital DMFT (IPT) approach we have calculated the transport properties. To check the angular dependence of MR we have applied magnetic field in different directions from \textbf{b-} to \textbf{a-} axis which can be easily done within WIEN2K code. See Supplementary Materials for details.\\

\section{Results}
\subsection{Resistivity and transverse magnetoresistance (TMR)}
\begin{figure}
\includegraphics[width=0.5\textwidth]{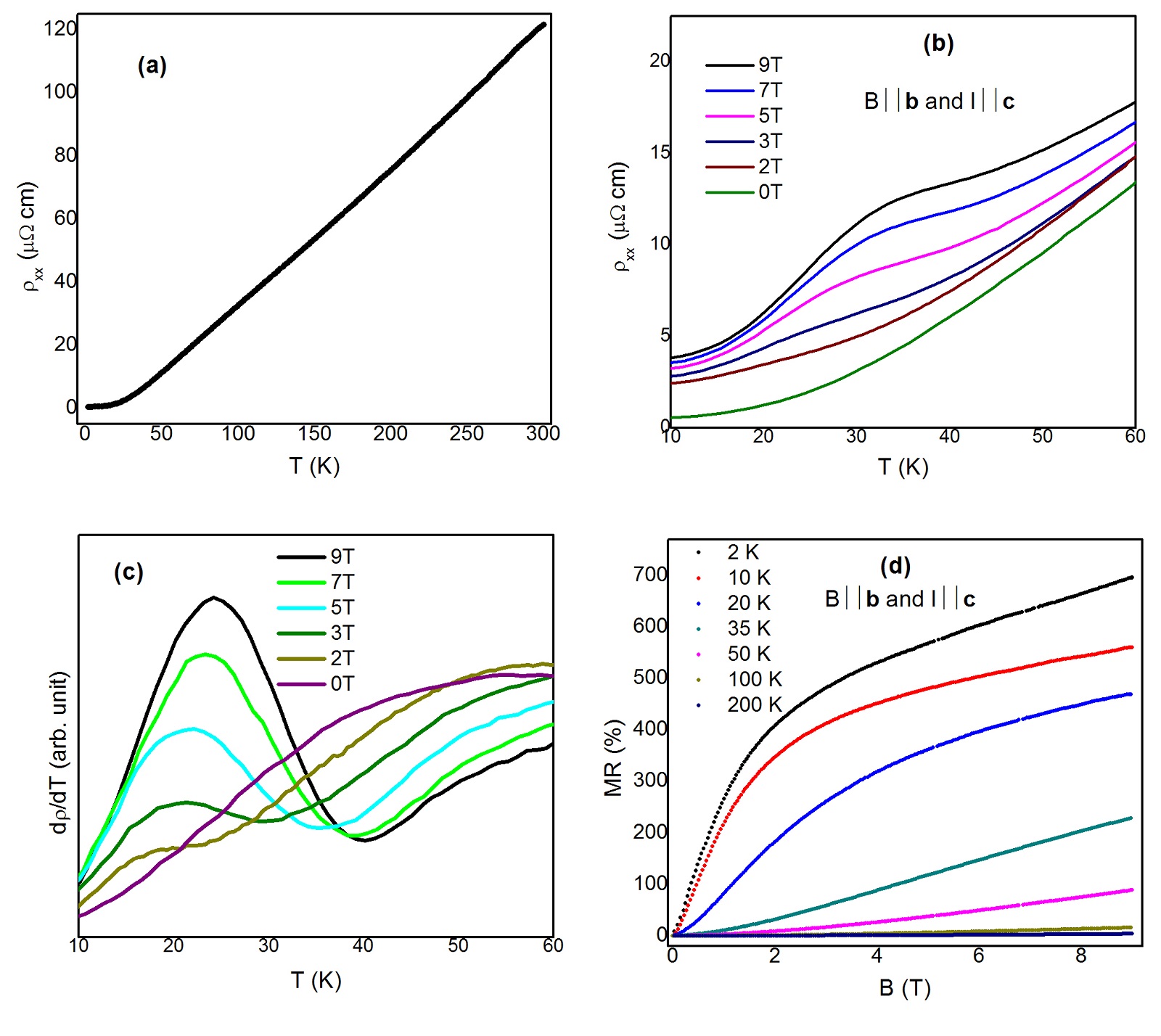}
\caption{Resistivity as a function of temperature and magnetic field in transverse configuration. (a) Temperature dependence of resistivity ($\rho_{xx}$) from 2 K to 300 K in absence of magnetic field ($B$). (b) $\rho_{xx} (T)$ up to 60 K at different external magnetic field strengths. (c) The first order derivative of $\rho_{xx} (T)$ to clearly locate the position of the hump in $\rho_{xx} (T)$. (d) TMR as a function of $B$ at some representative temperatures in $B$$\parallel$\textbf{b} and $I$$\parallel$\textbf{c} configuration.}\label{rh}
\end{figure}
The zero-field resistivity ($\rho_{xx}$) is metallic over the whole temperature range, as shown in Fig. 1a. $\rho_{xx}$ shows strong  linear-in-$T$ dependence at high temperature followed by a smooth crossover to $\rho_{xx}(T)\simeq \rho_{0}+AT^{2}$ below $T\simeq 20$ K, indicating the formation of Fermi liquid state. The coefficient $A$  is found to be very small ($A\sim$ 10$^{-6}$ $\mu\Omega cm/K^{2}$).  This suggests  electronic correlation in LaTe$_3$ is very weak. The small value of $\rho_{xx}$ at 2 K ($\sim$ 0.45 $\mu\Omega$ cm) and the large residual resistivity ratio (RRR), $\rho_{xx}$(300 K)/$\rho_{xx}$(2 K)$\sim$ 270, testify the high quality of LaTe$_{3}$ crystal. Similar metallic behaviour of $\rho_{xx}$ has also been reported with a RRR ($\sim$ 120) \cite{8,13}. One reason for the metallic behaviour of $\rho_{xx}$ below $T_{cdw}$ could be that the CDW state in LaTe$_{3}$ only partially gaps out the multi-sheeted Fermi surface due to imperfect nesting.  The temperature dependence of $\rho_{xx}$ becomes more exciting under application of magnetic field in $B$$\parallel$\textbf{b} and $I$$\parallel$\textbf{c} configuration, which has been shown in Fig. 1b. A hump-like feature gradually develops which shifts toward higher temperature with increasing field strength. The first-order derivative of $\rho_{xx} (T)$ in Fig. 1c clearly demonstrates that the position and sharpness of the anomaly depend strongly on the strength of magnetic field. In several magnetic RTe$_{3}$ compounds, similar hump-like feature in $\rho_{xx} (T)$ has been observed in absence of external magnetic field, which has been established as a characteristic signature of second charge density wave transition \cite{10,11,16,17}. Figure 1d shows the field dependence of the transverse magnetoresistance (TMR) in the same configuration, where MR is defined as $\frac{[\rho_{xx}(B) - \rho_{xx}(0)]}{\rho_{xx}(0)}\times 100\%$. At 2 K and 9 T, the value of TMR is as large as $\sim$ 700 \%  and it diminishes rapidly with increasing temperature. Interestingly, TMR increases monotonically with increasing $B$ without any sign of saturation up to 9 T. TMR exhibits two distinct regimes: a low-field quadratic $B$ dependence and a high-field linear $B$ dependence, connected by a smooth crossover.  With increasing $T$, the crossover shifts towards higher $B$. The crossover in TMR is more clearly visible in the $\frac{d(TMR)}{dB}$ versus $B$ plot, as shown in \textbf{Fig. S5a}. A non-saturating MR along with the linear-in-$B$ dependence is very interesting. This observation raises the enticing question about the nature of hitherto uninvestigated electronic state of LaTe$_{3}$ at high magnetic field.\\
\begin{figure}
\includegraphics[width=0.5\textwidth]{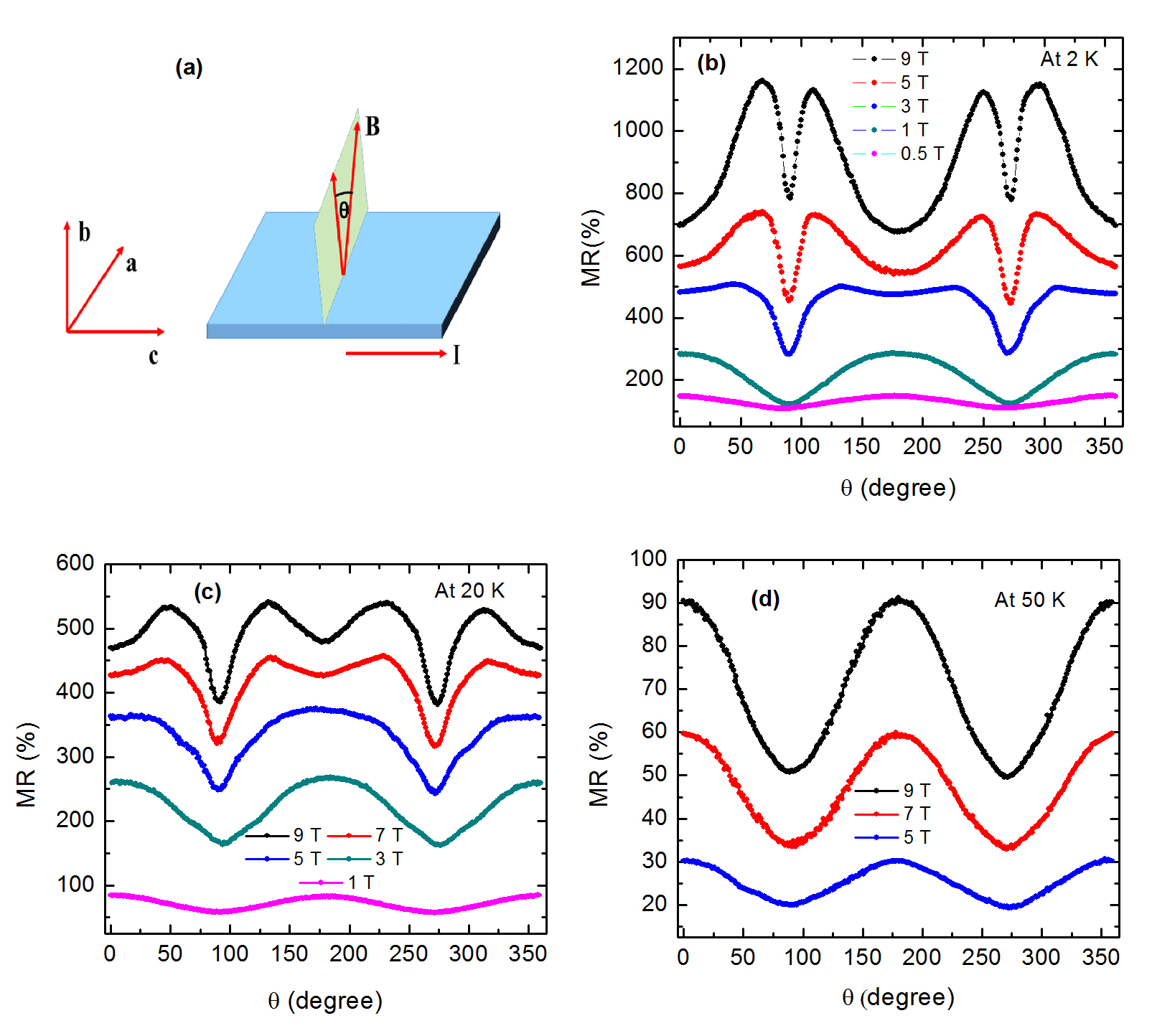}
\caption{Crystallographic direction dependence of magnetoresistance in transverse configuration. (a) A schematic diagram of the experimental configuration. The direction of magnetic field has been varied from \textbf{b}- to \textbf{a}-axis, making an angle $\theta$. (b), (c) and (d) are the $\theta$ dependence of TMR for different constant magnetic field strengths at 2 K, 20 K and 50 K, respectively.}\label{rh}
\end{figure}
Armed with the knowledge that the Fermi surface of LaTe$_{3}$ is diamond shaped in the \textbf{ac}-plane with minimal \textbf{b}-axis dispersion, we have measured TMR by rotating the field from \textbf{b}- to \textbf{a}-axis within \textbf{ab} plane. The schematic of experimental setup is shown in Fig. 2a. It is evident from Fig. 2b that TMR($\theta$) shows two-fold rotational symmetry at 2 K and 0.5 T, with minimum for $B$$\parallel$\textbf{a} ($\theta=90^{o}$) and maximum for $B$$\parallel$\textbf{b} ($\theta=0^{o}$) direction and $\frac{TMR(\theta=0)}{TMR(\theta=90^{o})}$ $\simeq$ 1.35.  With increasing $B$, initially this ratio increases and reaches $\sim$  2.3 at 1 T and then decreases monotonically to a value $\sim$ 0.88 at 9 T (\textbf{Fig. S5b in Supplementary Materials}). Thus, the low-field anisotropy in TMR reverses at high fields, a very unexpected characteristic that should ultimately have links to drastic modification of the electronic structure with $B$. Figures 2c and 2d show the angular variation of TMR at 20 K and 50 K, respectively, for different fields. One can argue from the figures that only the cos($\theta$) like  dependence in TMR, which has been observed at 2 K and low fields, survives for all the field strengths (up to 9 T) at high temperature above 20 K. The polar plots of TMR for different applied fields are shown in \textbf{Fig. S6} of Supplementary Materials. It is rewarding to make an analogy between the $B$ dependence of TMR in Fig. 1d and TMR($\theta$) at different field strengths in Figs. 2b-d, at a particular temperature. In the low-field region, where the TMR is quadratic in $B$, TMR($\theta$) shows cosine-like angular variation. On the other hand, at high fields, where TMR is linear in $B$,  TMR($\theta$) exhibits sine-like dependence.  This reveals strong causal connections between the quadratic-in-$B$ MR and cosine-like TMR($\theta$), and between the linear MR  and sine-like TMR($\theta$).  Thus, a tantalizing question arises: is there a field-induced Fermi surface reconstruction  with the high-$B$ state  exhibiting anomalies famously associated with semi-metallic behavior?

\subsection{Longitudinal magnetoresistance}
\begin{figure*}
\includegraphics[width=0.9\textwidth]{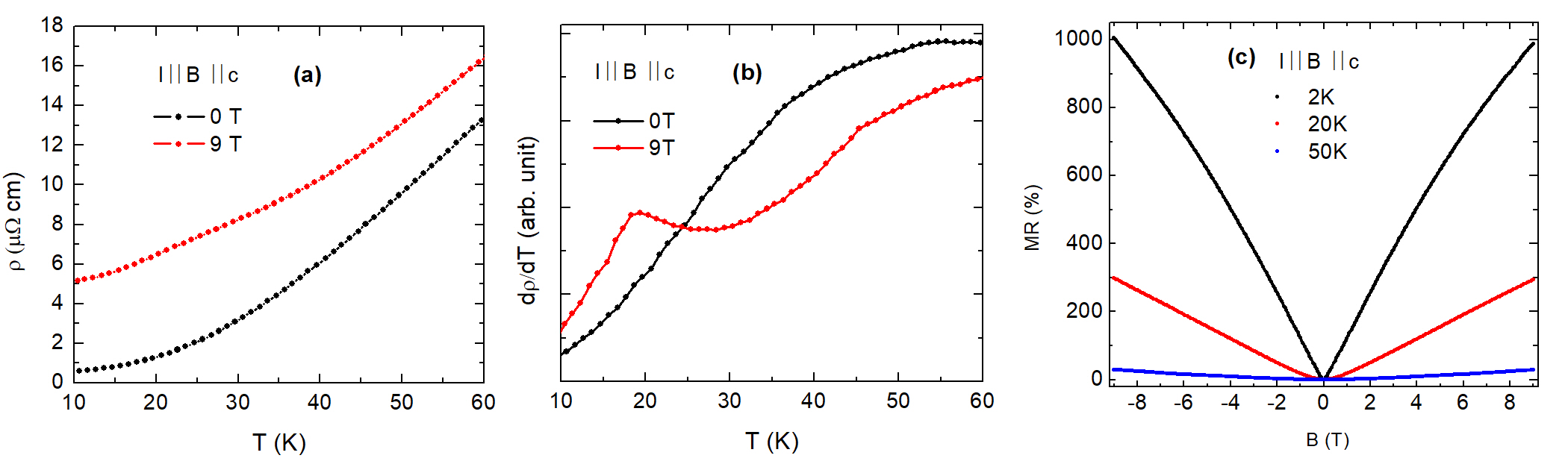}
\caption{Temperature and field dependence of magnetoresistance in longitudinal configuration when $B$$\parallel$$I$$\parallel$\textbf{c}. (a) Temperature dependence of resistivity ($\rho_{xx}$) at 0 and 9 T magnetic field ($B$). (b) The first order derivative of $\rho_{xx} (T)$ to clearly locate the position of the hump in $\rho_{xx} (T)$ under application of magnetic field. (a) Magnetic field dependence of LMR at some representative temperatures.}\label{rh}
\end{figure*}
Similar to transverse configuration, temperature dependence of resistivity under application of magnetic field has also been measured in longitudinal setup with $B$$\parallel$$I$$\parallel$\textbf{c} (Fig.3a). At 9 T, the hump-like feature can be observed in $\rho_{xx}(T)$ at around 28 K. Although the anomaly is weaker compared to that observed in Fig. 1b (transverse case), it is clearly visible in $\frac{d\rho}{dT}$ plot in Fig. 3b. Also, the hump appears at slightly lower temperature compared to that observed in transverse configuration. The magnetic field dependence of $\rho$ at low temperature may affect the position and nature of the anomaly. As shown in Fig.3c at 2 K and 9 T, LMR is $\sim$ 10$^{3}$ \%. LMR sharply rises with increasing $B$, undergoes a crossover from an approximate linear over a narrow range of field to sublinear B dependence at high fields and remains non-saturating up to 9 T. The much sharper increase in $\rho$ with $B$ for the longitudinal configuration weakens the anomaly and shifts its position toward lower temperature. Our theoretical calculations, to be discussed in the latter section, also support the appearance of a hump in $\rho_{xx}(T)$ curve for both transverse and longitudinal configuration.  Figures 4b-d show the angle dependence of MR at some representative temperatures for the experimental configuration shown in Fig. 4a. The direction of magnetic field has been varied from \textbf{b}- to \textbf{c}-axis, making an angle (90$^{\circ}$ - $\phi$) with the current direction, in the \textbf{bc} plane.  We uncover peculiar $\phi$ and $B$ dependence of MR, which has not been reported in any CDW system so far. As shown in Fig. 4b,  at 9 T and 2 K, the maximum value of MR occurs for $\phi$$=$90$^{\circ}$ (LMR). MR($\phi$) shows that two weaker spike-like peaks appear symmetrically around the strongest peak at $\phi$$=$90$^{\circ}$. The weakest (MR $\sim$ 590 \%) and intermediate (MR $\sim$ 770 \%) peaks in MR are $\sim$ 17$^{\circ}$ and $\sim$ 28$^{\circ}$ away from the strongest one, respectively. With decreasing $B$, the height of the prominent peak at $\phi$$=$90$^{\circ}$ reduces drastically, while the other peaks show higher immunity to field. As a result,  MR($\phi$) exhibits a minimum at $\phi$$=$90$^{\circ}$ for $B$$<$3 T.  Figures 4c-d show that the peak at $\phi$$=$90$^{\circ}$ also becomes very weak with increasing $T$. At high temperature, the nature of MR($\phi$) curve is to some extent similar to that observed at 2 K for low field strengths.
\begin{figure*}
\includegraphics[width=0.9\textwidth]{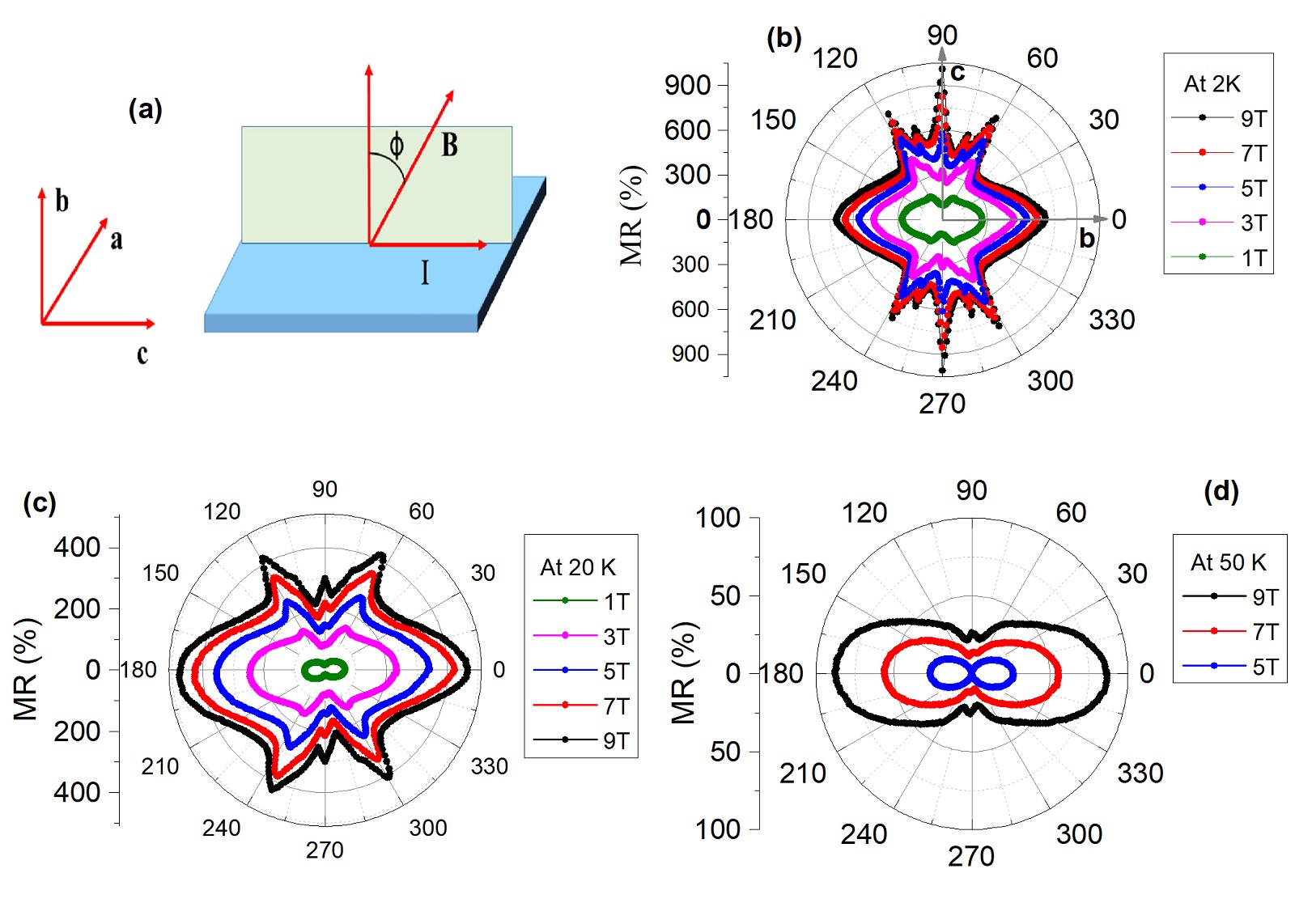}
\caption{The dependence of magnetoresistance on the angle between $I$ and $B$ in bc-plane of the crystal. (a) Schematic diagram, illustrating the experimental configuration to measure the angular variation of MR in \textbf{bc}-plane. The direction of magnetic field has been varied from \textbf{b}- to \textbf{c}-axis, making an angle (90$^{\circ}$ - $\phi$) with the current direction. The polar plot of MR($\phi$) (b) at 2 K, (c) at 20 K, and (d) at 50 K, respectively, for some representative magnetic field strengths.}\label{rh}
\end{figure*}
\section{Temperature and field dependence of Hall conductivity}
\begin{figure*}
\includegraphics[width=0.9\textwidth]{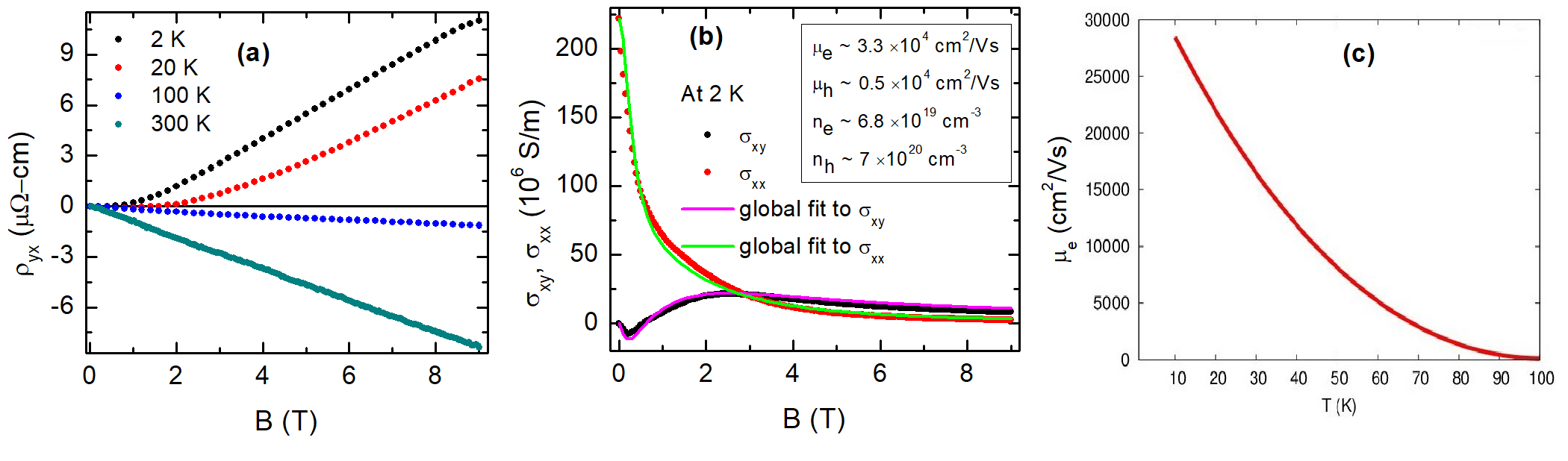}
\caption{Hall measurements and multi-band analysis. (a) Field dependence of the Hall resistivity ($\rho_{yx}$) at different temperatures. (b) Global two-band fitting of the Hall conductivity ($\sigma_{xy}$=$\frac{\rho_{yx}}{\rho_{yx}^{2} + \rho_{xx}^{2}}$) and electrical conductivity ($\sigma_{xx}=\frac{\rho_{xx}}{\rho_{yx}^{2} + \rho_{xx}^{2}}$) at a representative temperature 2 K. (c) Theoretically estimated electron mobility from transport calculation and its temperature dependence.}\label{rh}
\end{figure*}
To determine the nature and density of the charge carriers, the Hall resistivity ($\rho_{yx}$) as a function of $B$ has been measured in the temperature range 2-300 K, shown in Fig. 5a. At 300 K, $\rho_{yx}$ is found to be negative and almost linear in $B$, indicating electron dominated transport near room temperature. Upon reducing $T$, $\rho_{yx}$($B$) becomes nonlinear and changes sign at high fields. This implies the presence of both the electron and hole-type charge carriers, and their contributions to electronic transport are comparable to each other. We have performed simultaneous fitting (i.e., Global fitting) of Hall conductivity ($\sigma_{xy}$=$\frac{\rho_{yx}}{\rho_{yx}^{2} + \rho_{xx}^{2}}$) and electrical conductivity ($\sigma_{xx}$=$\frac{\rho_{xx}}{\rho_{yx}^{2} + \rho_{xx}^{2}}$) data with the expressions derived from the semiclassical two-band model \cite{22}. Figure 5b shows the theoretical fit to $\sigma_{xy}$ and $\sigma_{xx}$ with the expressions,
\begin{equation}
\sigma_{xy}(B)=\left[\frac{n_{h}\mu_{h}^{2}}{1+(\mu_{h}B)^2} - \frac{n_{e}\mu_{e}^{2}}{1+(\mu_{e}B)^2}\right]eB,
\end{equation}
and
\begin{equation}
\sigma_{xx}(B)=\left[\frac{en_{h}\mu_{h}}{1+(\mu_{h}B)^{2}} + \frac{\sigma_{xx}(0) - en_{h}\mu_{h}}{1+(\mu_{e}B)^{2}}\right].
\end{equation}
$n_{e}$ ($n_{h}$) and $\mu_{e}$ ($\mu_{h}$) are electron (hole) density and electron (hole) mobility, respectively. The obtained values of $n_{e}$ and $n_{h}$ are $\sim$ 6.8$\times$10$^{19}$  and 7.0$\times$10$^{20}$ cm$^{-3}$, respectively. On the other hand, $\mu_{e}$ ($\sim$3.3$\times$10$^{4}$ cm$^{2}$ V$^{-1}$ s$^{-1}$ at 2 K) is found to be significantly higher than $\mu_{h}$ ($\sim$0.5$\times$10$^{4}$ cm$^{2}$ V$^{-1}$ s$^{-1}$). Such large values of carrier mobility have not been reported earlier in any rare-earth tritellurides, and are also rare in CDW and vdW compounds. Interestingly, the value of carrier mobility is comparable to that recently observed in topological semimetals \cite{23,24,25,26}. The density of charge carriers has also been estimated from the band structure theory calculation. These values, n$_e$=6.9 $\times$ 10$^{19}$ and n$_h$=7.2 $\times$ 10$^{20}$ cm$^{-3}$ are close to that obtained from the two-band fitting of electrical and Hall conductivity. Not only the carrier density, the electron mobility has also been estimated from transport calculations and shown in Fig. 5c. The calculated mobility is close to that obtained from two-band fitting.

\section{Discussions}
\begin{figure}
\includegraphics[width=0.5\textwidth]{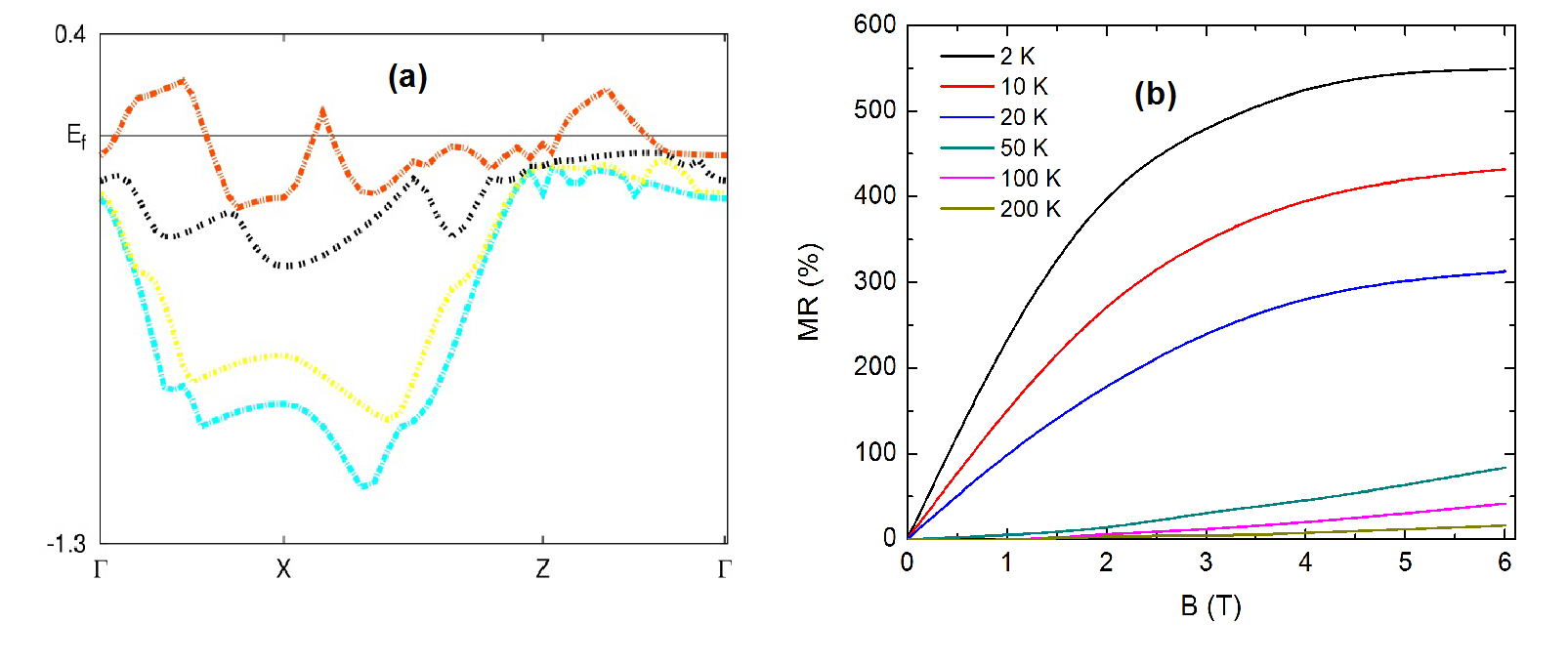}
\caption{Electronic band structure, and field dependence of TMR from transport calculations. (a) Electronic band structure of LaTe$_{3}$ in the range -1.3 to 0.5 eV about the Fermi level. The Fermi level crossing band is from Te $p$ orbital. (b) TMR as a function of external magnetic field up to 6 T, at some representative temperatures and in $I$$\parallel$$c$-axis and $B$$\parallel$$b$-axis configuration.}\label{rh}
\end{figure}
Transverse magnetoresistance in a non-magnetic compound is typically orbital in nature and it scales with the mobility ($\mu$) of charge carriers in the plane perpendicular to the applied magnetic field \cite{27,28}. For example, considering the contribution of one type of charge carrier with density ($n$), the typical field dependence of electrical conductivity is given by $\sigma(B)$ $\sim\frac{ne\mu}{1+\mu^2 B^2}$ \cite{27,28}. So materials having high carrier mobility generally show large MR. Famously, compensation or near-compensation of electrons and holes also results in large and non-saturating transverse MR: instances would be elemental bismuth \cite{29} and possibly some recently discovered topological semimetals \cite{23,30}. From the Hall measurements, band-structure theory and transport calculations, it is evident that the mobility of charge carriers in LaTe$_{3}$ is very large and the contribution of electrons and holes to transport is comparable. This might be the possible explanation for the large and non-saturating MR in LaTe$_{3}$.

From first-principles density-functional theory-plus dynamical mean-field theory (DFT+DMFT) and transport (involving the full DFT+DMFT propagators without irreducible vertex corrections) calculations, we obtain the electronic band structure (Fig. 6a) as well as the field dependence of MR for $I$$\parallel$$c$-axis and $B$$\parallel$$b$-axis configuration (Fig. 6b).  In accord with previous work, the $p$ band originating from planar Te atoms generates the Fermi surface \cite{13}. Remarkably, the theoretically obtained field dependence of MR is very similar to that shown in Fig. 1d in all major respects.  Moreover, the computed value of MR is also comparable to that observed in experiment. Several interesting features, germane to the magnetotransport findings, stand out from our DFT+DMFT studies:\\

$(i)$ At zero field, the DFT Fermi surface (Fig. 7a) agrees with earlier band structure calculations and ARPES measurements \cite{12,13}.  On the other hand, the Fermi surface for finite $B$ (both in- and out-of-plane configuration, as shown in Fig. 7b and Fig. 7c, respectively) exhibits a reduction of in-plane curvature, suggesting gradual evolution toward more efficient nesting.  This is very clear in Fig. 7c, (in an in-plane field $B=3.0$ T) where the reconstruction is particularly drastic:  the Fermi surface now has large flat portions, almost perfectly nested with each other. Interestingly, nesting tendency even between different Fermi sheets emerges at $B=3.0$ T.  Further, since the inner closed Fermi surface sheet arises predominantly from the Te $p_{x}$ and $p_{z}$ orbitals, the field-induced ``flattening" of this sheet in an in-plane field $B$=3.0 T  actually suggests emergence of two quasi-one-dimensional bands, weakly hybridized with each other.  Though this effect is weaker for out-of-plane fields, the reduction of in-plane Fermi surface curvature is clearly visible in that case as well.  Enhancement of flat regions in the Fermi surface immediately implies enhancement of scattering due to electron-electron and electron-phonon interactions, providing a direct insight into large and positive MR.\\

$(ii)$  Both, electronic correlations and electron-lattice coupling renormalize the band dispersions and the DFT Fermi surface(s).  Within local dynamical mean-field theory, a ${\bf k}$-independent but energy ($\omega$)-dependent self-energy generically reduces the DFT band widths, but leaves the shape {\it and} size of the Fermi surface nearly unaffected.  However, in multi-band systems, interactions can also lead to band-dependent renormalization of electronic states. We have carried out DFT+DMFT calculations by varying the intra-orbital ($U$) and inter-orbital ($U'$) interaction in the range $0.0 < U \leq 2.0$, and $U'=0.3U$.  For $U=1.0$ eV,  self-energies for both $p_{x}$ and $p_{z}$ carriers exhibit a Landau-Fermi liquid form up to rather high $T$, $i.e$, Im$\Sigma_{p_{x},p_{z}}(\omega) \simeq -a\omega^{2}$ at low energy.  Interestingly, for $U=2.0$ eV, we uncover a coherence-incoherence crossover as a function of $T$: Im$\Sigma_{p_{x},p_{z}}(\omega) \simeq -a'\omega^{2}$ for $T<T_{coh}\simeq 50$ K smoothly crosses over to $\simeq -A-a''\omega^{2}$ for $T>T_{coh}$.  Correspondingly,  the dc resistivity crosses over from a quasilinear-in-$T$ to a $T^{2}$ dependence  at low temperature, in nice qualitative accord with experiment.  Even more interestingly, in a magnetic field $B=1$ T, and for $U=2.0$ eV, Im$\Sigma_{p_{z}}(\omega)$ exhibits a (negative) pole-like structure.  This directly implies a rapid enhancement of the scattering rate in a magnetic field, and can be rationalized in terms of a reduction of the DFT bandwidth ($W$) for $B>0$, leading to stronger inelastic scattering and providing a possible explanation of positive MR.  Remarkably, we find that both, the dc resistivity as well as the sign and magnitude of the MR agree quite well with experimental data, providing good support to this interpretation. It is interesting to note that the magnetic field-induced hump-like feature in $\rho$($T$) curve appears close to the coherence-incoherence transition temperature.\\

$(iii)$  The emergence of a renormalized electronic structure comprising two ``crossed'' quasi-one-dimensional bands is quite interesting. Even a slight interaction on the carriers in two one-dimensional ($1D$), $p_{x}$- and $p_{z}$-derived, bands may have strong effects.  It is interesting that intermediate-energy power-law frequency dependence of the optical conductivity in RTe$_{3}$ systems has been observed \cite{31}, though the electronic origin of quasi-$1D$ bands necessary for this picture has hitherto remained unclear here.  An interchain $p_{x}-p_{z}$ hybridization favors Landau Fermi-liquid coherence, and the interband electron-electron (electron-phonon) interactions, in turn, could be susceptible to instability to an unusual CDW state.\\

$(iv)$ Our theoretical approach also gives very good agreement with details of both, longitudinal {\it and} transverse MR as functions of in-plane and out-of-plane magnetic field.  The good accord with the $T$-dependent resistivity as well as the sign, magnitude and non-saturating nature of MR at high field (at least up to $9$ T), permits a consistent interpretation of $B$, $T$ and angle-dependent MR in terms of a field-dependent Fermi surface reconstruction (a field-induced Lifshitz transition) {\it inside} the CDW phase.  We are unaware of such behavior, {\it viz}, a field-induced Lifshitz transition and the coherence-incoherence crossover in resistivity occurring well below $T_{cdw}$, in transition-metal and rare-earth tri-chalcogenides.\\
\begin{figure*}
\includegraphics[width=0.9\textwidth]{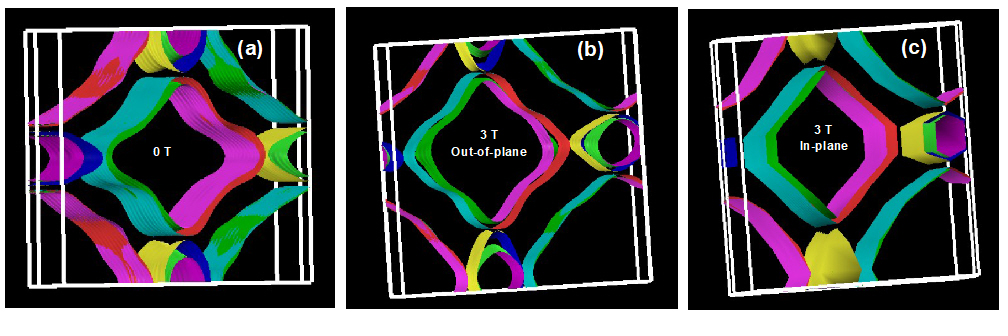}
\caption{Theoretically simulated Fermi surface in presence and absence of external magnetic field. Fermi surface of LaTe$_{3}$ (a) at zero field, (b) at 3 T out-of-plane, and (c) at 3 T in-plane magnetic field.}\label{rh}
\end{figure*}
In non-magnetic materials, LMR is usually found to be negligible. Although one could conceive of several alternative scenarios, they fall short of providing a unified rationalization of the above observations. A more plausible scenario could be the following. In a few ultra-clean metals, the LMR can be large and positive, and tends to saturate at high fields \cite{32,33}. It has been ascribed to the multiply connected nature of the Fermi surface through Brillouin zone boundaries. When the cross-sectional area of the necks is large compared to the total area of the Fermi surface, the conductivity experiences strong suppression in external magnetic field.  The  calculated Fermi surfaces, shown in Figs. 7a-c and \textbf{Fig. S7}, are indeed  multiply connected across the Brillouin zone boundaries. While one could ascribe large and positive LMR to this aspect, the lack of saturation at high field is not consistent with the above mechanism.\\

However, the sizable field-induced FS reconstruction we find affords an explanation of the MR data.  We notice that similarities between the observed TMR of LaTe$_{3}$ and those of other materials with spin- and charge-density-wave states \cite{34} suggest that the crossover from $\Delta\rho \sim B^{2}$ to $\Delta\rho \sim |B|$ with increasing $B$ maybe intrinsic to field-driven  changes in FS as follows: the FS undergoes a clear ``topological'' change, from a smoothly curved at small $B$, to $(i)$ pockets with very flat, parallel sheets, and more importantly, to $(ii)$ several small pockets with sharp corners. Carrier motion around these sharp corners of FS will dominate the MR (over contributions  from flat parts of the FS), because large enhancement of the cyclotron frequency, $\omega_{c}$, at the corners will sizably enhance $\omega_{c}\tau$ \cite{35}.  Pippard showed that a square FS with infinitely sharp corners gives $\Delta\rho \sim |B|$, a behavior persisting as long as the mean free path $l_{k}$ of the charge carrier, is much larger than $r_{k}$, the cyclotron radius at the sharp corner \cite{35}.  A perusal of our Fig. 7 clearly shows that the large, linear MR is intimately correlated with the emergence of such a field-reconstructed Fermi surface.\\

Thus, the crossover from $B^{2}$- to a $|B|$-variation of the MR is intricately tied to the drastic modification of FS topology in modest magnetic fields.  Within our modelling, this can be understood as a manifestation of the interplay between the ${\bf k}$-dependent inter-band hybridization ($V({\bf k})$) and spin-orbit coupling, SOC ($\lambda$).  While the former is of order 200 meV, sizably larger than the latter, about 20 meV at the bare level, even small local Coulomb interactions could change this picture: $V$ is renormalized downward to $V_{ren}({\bf k})=z_{p}\, V({\bf k}) \simeq 50$~meV, ($z_{p} \simeq 0.25$ is the quasiparticle weight of the p-band) and the SOC, being a local quantity, is expected to be mildly enhanced by local correlations.  Thus, when $\lambda/V_{ren}({\bf k})$ is not small (about 0.4), considerable k-space rearrangement of electronic states occurs due to interplay between the Zeeman energy ($g\mu_{B}B$), $\lambda$ and $V_{ren}({\bf k})$: the coherent interband hybridization is sizably suppressed as a result, directly leading to destruction of FS warping features and emergence of flat parallel portions with sharp edges in Fig. 7.  This is the underlying mechanism for the striking sensitivity of the FS to a modest magnetic field we find in DFT+DMFT.\\

\begin{figure*}
\includegraphics[width=0.9\textwidth]{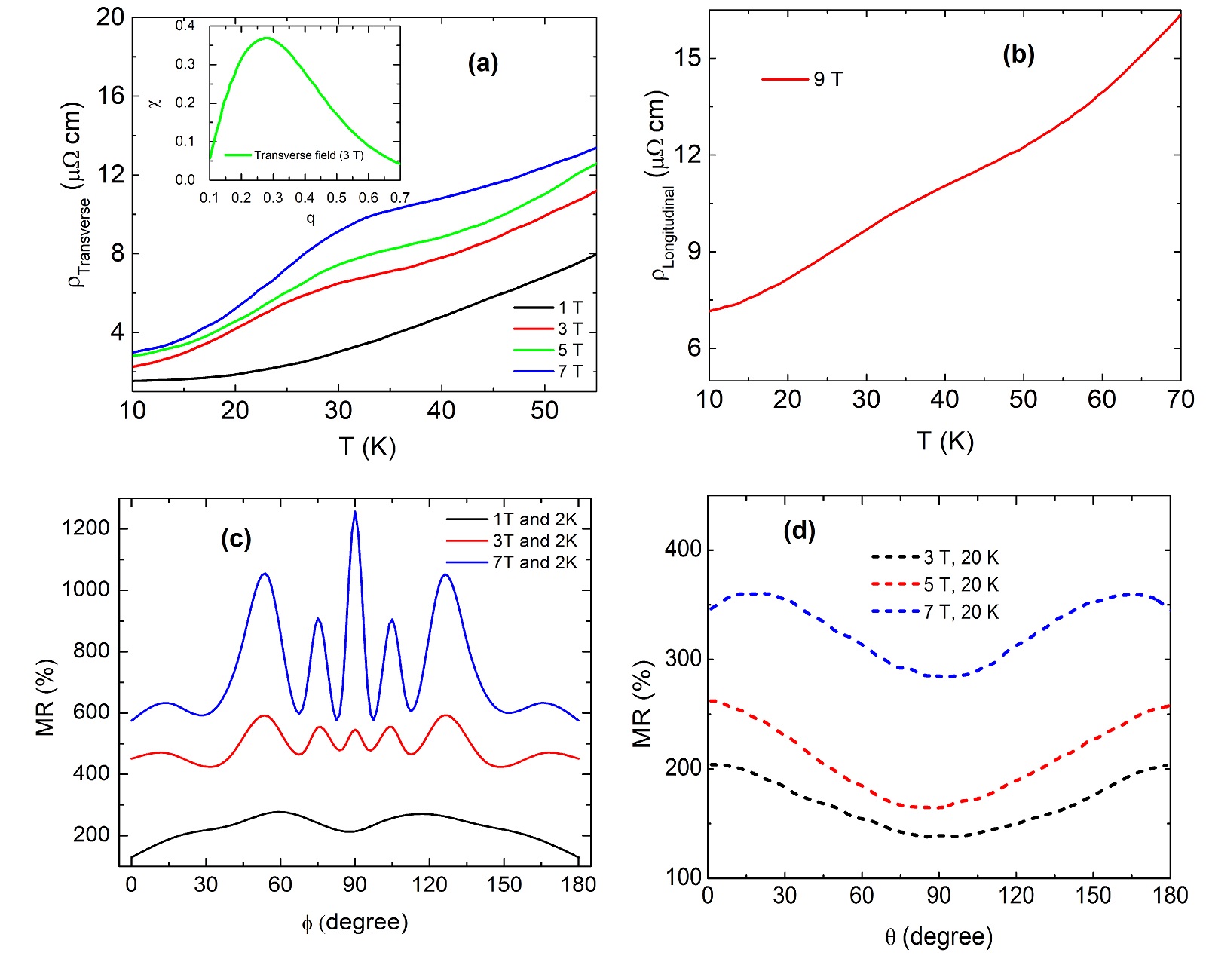}
\caption{Temperature dependence of resistivity, theoretically obtained susceptibility plots, and the angle dependence of MR obtained from transport calculations. Resistivity as a function of temperature both in absence and presence of $B$ in the: (a) $I$$\parallel$$c$-axis and $B$$\parallel$$b$-axis configuration as in Figure 1b and (b) $B$$\parallel$$I$$\parallel$\textbf{c}-axis configuration as in Figure 3a. he susceptibility plot, obtained at representative field strength 3 T for transverse field configuration has been shown in the Inset of Fig. 8(a). The $\chi$-plot which has been obtained at 30 K represents the possible low-temperature CDW with $Q=1/3a* $.(c) MR as a function of angle between current and magnetic field, in the bc-plane of the crystal. (d) Crystallographic direction dependence of TMR which has been observed by rotating the direction of magnetic field about current parallel to c-axis.}\label{rh}
\end{figure*}

Moreover, the anomalous behavior of MR($\phi$), wherein spike-like peaks at the intermediate angles appear when the direction of magnetic field is changed from $B$$\parallel$\textbf{b} to $B$$\parallel$\textbf{c}$\parallel$$I$ configuration, testifies to a geometrical effect due to quasi-two-dimensional nature of the Fermi surface(s) \cite{32,36,37,38}. In high-purity layered metallic systems, at some characteristic angles, known as Yamaji angles, the cyclotron orbits on the corrugated Fermi surface acquire equal cross-sectional areas. As a consequence, the Landau orbits become non-dispersive and the group velocity of electrons perpendicular to the layers vanishes, leading to a peak in the resistivity. In the present material, MR($\phi$) shows strong Yamaji peaks at $\sim$ 73$^{\circ}$ and $\sim$ 62$^{\circ}$. This clearly shows the significant contribution of the CDW reconstructed Fermi surfaces to the geometrical effect in MR($\phi$). The suppression of Yamaji peaks with increasing temperature may result from $(i)$ thermal broadening of Landau orbital states in a pure one-electron band-structure view, or $(ii)$ in a correlated view, the coherence-incoherence (C-IC) crossover in LaTe$_{3}$ occurs at $T_{coh}\simeq 50$ K for $B=0$, which results in loss of Landau quasiparticles themselves. Since the Fermi surfaces show marked propensity towards development of flatter sections for $B>0$, one expects a reduction of the one-electron band-width, and hence of the C-IC crossover scale with $B$, leading to damping out of the Fermi surfaces themselves above $T_{coh}(B)$.\\

As mentioned earlier, in the absence of a magnetic field, the magnetic rare-earth tritellurides exhibit a second CDW transition ($T_{CDW2}$) at much lower temperature in a perpendicular direction to the higher temperature one, around the wave vector $Q = 1/3a*$.  The non-magnetic tritellurides, though, do not show this low-temperature CDW transition \cite{4,10,11,12,16,17}. Temperature dependence of resistivity shows a weak  anomaly at $T_{CDW2}$. In LaTe$_3$,  the appearance of a hump-like feature in $\rho$ ($T$) curve with applied field and the systematic enhancement of its sharpness with increasing field strength appear to suggest a connection of this anomaly to a field-induced CDW transition. To probe this connection, we have performed a charge-susceptibility calculation at 30 K in the presence of 3 T external magnetic field.  The susceptibility plot in the inset of Figure 8(a) (at transverse field configuration as representative), shows a peak at $Q = 1/3a*$. This suggests that the magnetic field is likely to induce an additional low temperature CDW state in the non-magnetic LaTe$_3$ at the same value of Q observed in magnetic RTe$_3$ compounds (in the absence of magnetic field). Similar results have also been obtained for longitudinal configuration in our susceptibility calculation. Considering the propensity of the field-reconstructed Fermi surface towards enhanced nesting, the low temperature hump in resistivity-temperature curves (Figs. 1b,c and Figs. 3a,b) has been successfully reproduced from theoretical calculations, shown in Figs. 8a,b in transverse and longitudinal configurations. The position of the hump, its variation with field strength, and its intensity over background resistivity are in good agreement with experimental findings.\\

The theoretically computed MR($\phi$) (Fig. 8c) shows fairly good agreement with experimental results (Fig. 4b). Both, the number of MR peaks and their positions are successfully reproduced by theory. Furthermore, the magnitudes of the MR in both the cases at a given angle are also comparable to experimental values of MR.  Similar to MR($\phi$), the experimental MR($\theta$), as shown in Fig. 2b, is also well reproduced by theory (see Fig. 8d). Not only the angle dependence and the value of MR but also the extreme sensitivity of the angular variation of MR($\phi$) and MR($\theta$) on  field is reproduced by theory, as evident in Fig. 8c-d. This testifies to the intimate link of these novel features with the field-induced modification of Fermi surface, as shown in Fig. 7a-c.  Taken together, these findings support a field-induced Lifshitz transition of the zero-field FS.
\section{Concluding remarks}
In conclusion, the present research demonstrates a hitherto uninvestigated through electronic transport experiments on a candidate of rare-earth tritellurides, LaTe$_{3}$. We observe several intriguing phenomena in the magnetotransport properties, rarely seen in any CDW system. A novel manifestations of the interplay between CDW order and field-induced electronic structure modification in a non-magnetic material has been exposed. The study holds promising potential to establish new route to tune CDW states. In addition, Hall measurements supported by our transport calculation reveals an unexpected high mobility of charge carriers, which has remained elusive for these CDW and vdW materials. The unique coexistence of weak inter-layer van der Waals coupling, CDW state, interplay of CDW with external perturbation like magnetic field, and huge mobility of charge carriers, may have promising impact on technological applications and device fabrication.
\section{Acknowledgments}
We acknowledge Prof. Sudipta Roy Barman and his group, UGC-DAE Consortium for Scientific Research, Indore for LEED measurements. S. K acknowledge the support from DST women scientist grant SR/WOS-A/PM-80/2016(G).
\newpage
\pagebreak

\newpage
\pagebreak

\begin{center}
\textbf{Supplementary material for "Fascinating interplay between charge density wave instability and magnetic field in rare-earth tritellurides, LaTe$_{3}$"}
\end{center}
\section{Details of First-principles and transport calculations}
We combine density functional theory (DFT) and dynamical mean field theory (DMFT) in order to examine the electronic structure and resulting properties of LaTe$_{3}$. Calculations are based on the experimentally determined Cmcm structure and lattice parameters, as described in the present and earlier studies \cite{1,2}. DFT calculations for LaTe$_{3}$ have been performed using the WIEN2k full-potential linearized augmented plane wave (FP-LAPW) ab initio package \cite{3,4}. For the calculations, 1000 k-points (10 $\times$ 10 $\times$ 10 mesh), cutoff parameter Rk$_{max}$= 7.5 (R is the smallest muffin-tin radius and k$_{max}$ is the cut-off wave vector of the plane-wave basis set), and generalized gradient approximation Perdew–Burke–Ernzerhof (GGA-PBE) exchange–correlation potential were chosen. The muffin-tin radius (R) [a.u.] was set to 2.5. These parameters are converged such that energy convergence are accurate to 0.0001 eV in the irreducible Brillouin zone. Calculations are carried out in both without and with magnetic field with inclusion of spin-orbit coupling. To include the effect of local correlations DMFT+IPT calculations were carried out with input band structure from WIEN2K. The energy range for the calculations were taken from -10 eV to 10eV to capture the higher energy contribution of the Te5p orbitals to the density of states arising from the hybridization in between p$_x$ and p$_z$ orbitals. Te p$_x$ and p$_z$ orbitals therefore constitute 2 $\times$ 2 Hamiltonian, which effectively results into two bands. The DMFT impurity problem was solved using the iterated perturbation theory approach. The multi orbital iterated perturbation theory (MO-IPT) is used as an impurity solver in DMFT: though not exact, it is a computationally fast and effective solver, and has been proven to work very well in real multi-band systems throughout all temperature range \cite{5,6,7}. We first carried out band structure calculation obtained from DFT. We observe one band is crossing Fermi level and forms electron like Fermi surface sheets around X point (Fig. 6). With applied magnetic field two more bands cross Fermi level. These bands are mainly from Te-p orbital. Two-dimensional cuts through the Fermi surface (FS) were extracted from WIEN2k on 200 $\times$ 200 k point grids for the LDA calculations. The Fermi surface was interpolated linearly and shown in Fig. 7. Comparing these DFT results to earlier FS measurements \cite{2} we find that the agreement in size and shape along the high symmetry directions is quite well. The FS in the parent state consists of inner and outer diamond sheets which are formed from Te $p_x$ and $p_z$ orbital. With zero magnetic field FS is symmetric along $k_x$ and $k_z$ direction but the symmetry is destroyed once the magnetic field is switched on. Correlation is induced in the Te-3p orbitals via DMFT, which changes the noninteracting electronic structure significantly. In the main text, we present our LDA+DMFT results. Using Multi-Orbital DMFT (IPT) approach we have calculated the transport properties (Fig. 6b and Fig. 8). To check the angular dependence of MR we have applied magnetic field in different directions from \textbf{b}- to \textbf{a}- axis which can be easily done within WIEN2K code.

\section{X-ray diffraction on single crystal sample and crystal structure image}
X-ray diffraction (XRD) was performed on the single crystal sample. As shown in Fig. S1a, the presence of very sharp (0 k 0) peaks in diffraction pattern confirms that the flat plane of the crystal is perpendicular to the crystallographic \textbf{b}-axis.\\
\begin{figure*}[h]
\includegraphics[width=0.8\textwidth]{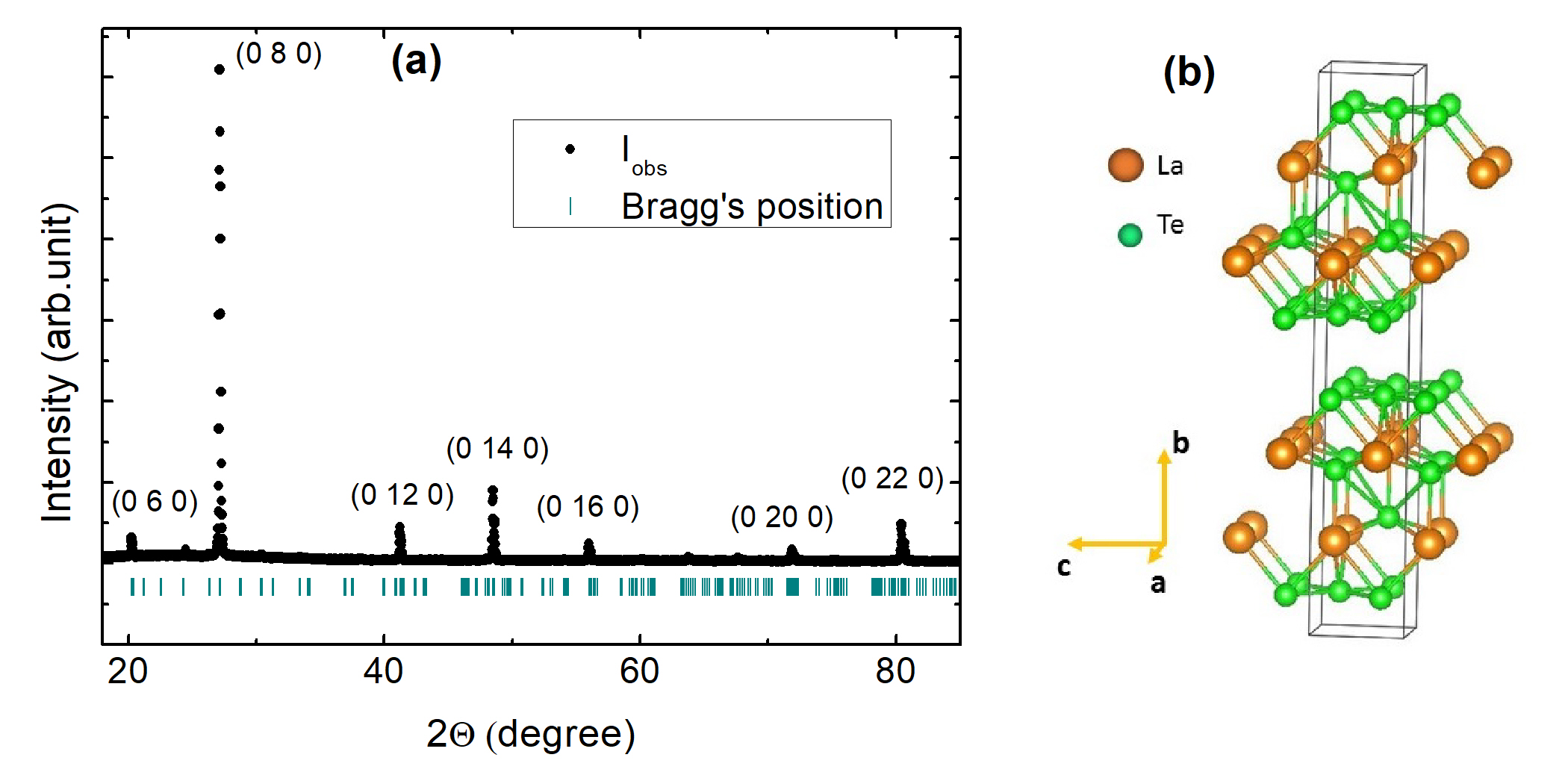}
\caption*{\textbf{Figure S1 $\mid$ X-ray diffraction on single crystal sample and crystal structure of LaTe$_{3}$.} (a) X-ray diffraction from the largest flat surface of the crystal. Black circles are experimental data (I$_{obs}$) and blue vertical lines show the position of Bragg's peaks. (b) Crystal structure image of LaTe$_{3}$, generated by VESTA software, using lattice parameters as input.}
\end{figure*}

\section{Powder x-ray diffraction and profile refinement analysis}
Figure S2 shows the high-resolution x-ray diffraction pattern of the powdered sample of LaTe$_{3}$ crystals at room temperature. Within the resolution of XRD, we did not see any peak due to the impurity phase. Using the Rietveld profile refinement, all the peaks in the diffraction pattern can be indexed with a orthorhombic unit cell (space group
Cmcm) having $a$=4.392(2), $b$$=$26.245(6), and $c$=4.417(3) {\AA}. The obtained values of lattice parameters are consistent with earlier reports \cite{1,2,8}.
\begin{figure*}[h]
\includegraphics[width=0.8\textwidth]{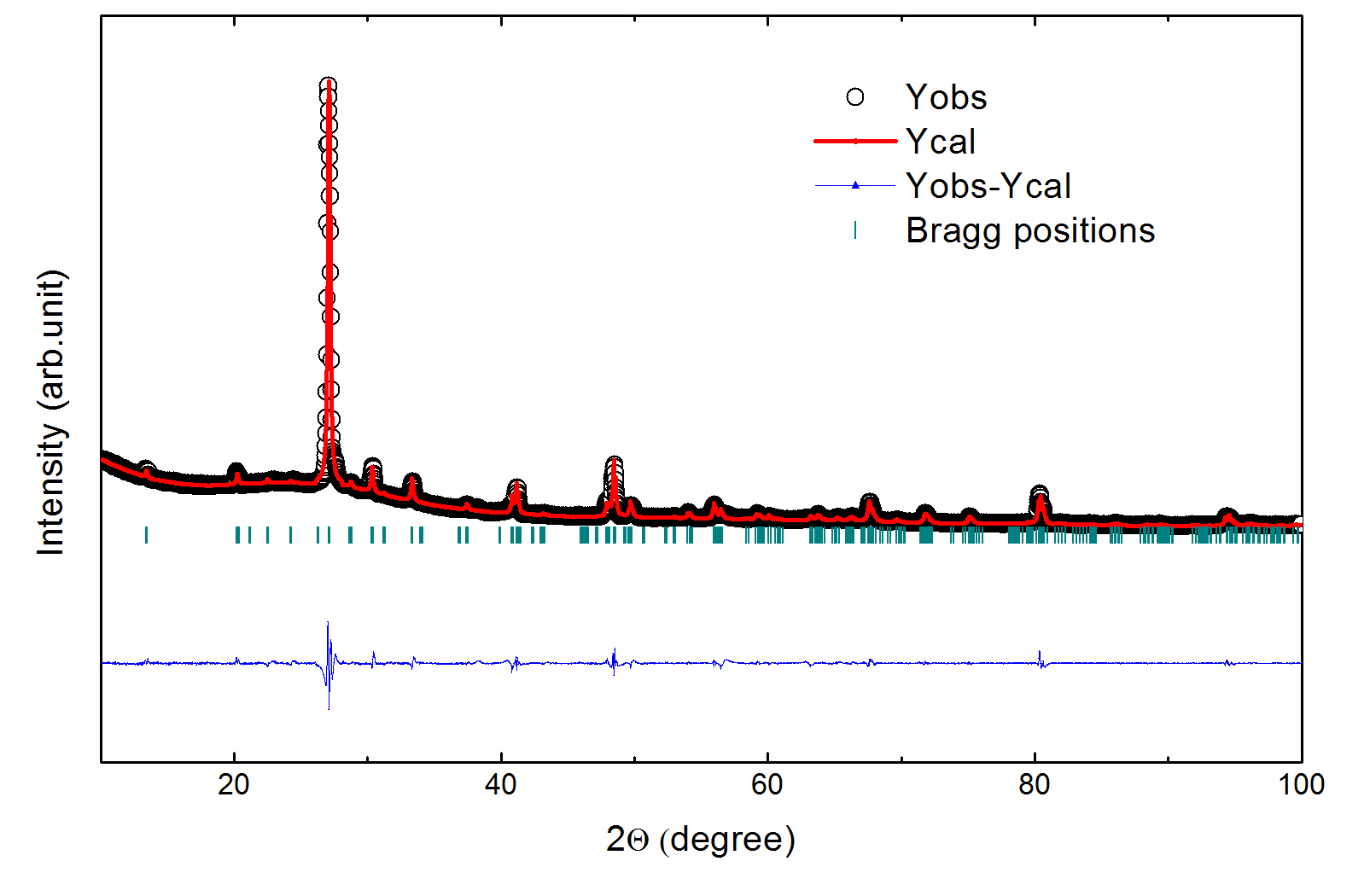}
\caption*{\textbf{Figure S2 $\mid$ Powder x-ray diffraction on crashed single crystals.} Black open circles are experimental data (Y$_{obs}$), red line is the calculated pattern (Y$_{cal}$), blue line is the difference between experimental and calculated intensities (Y$_{obs}$-Y$_{cal}$), and green lines show the Bragg positions.}\label{rh}
\end{figure*}

\section{Low-Energy electron diffraction (LEED) study of as-grown LaTe$_{3}$ crystals}
Result of our LEED measurements has been shown and illustrated in Figure S3. 
\begin{figure*}[h]
\includegraphics[width=0.8\textwidth]{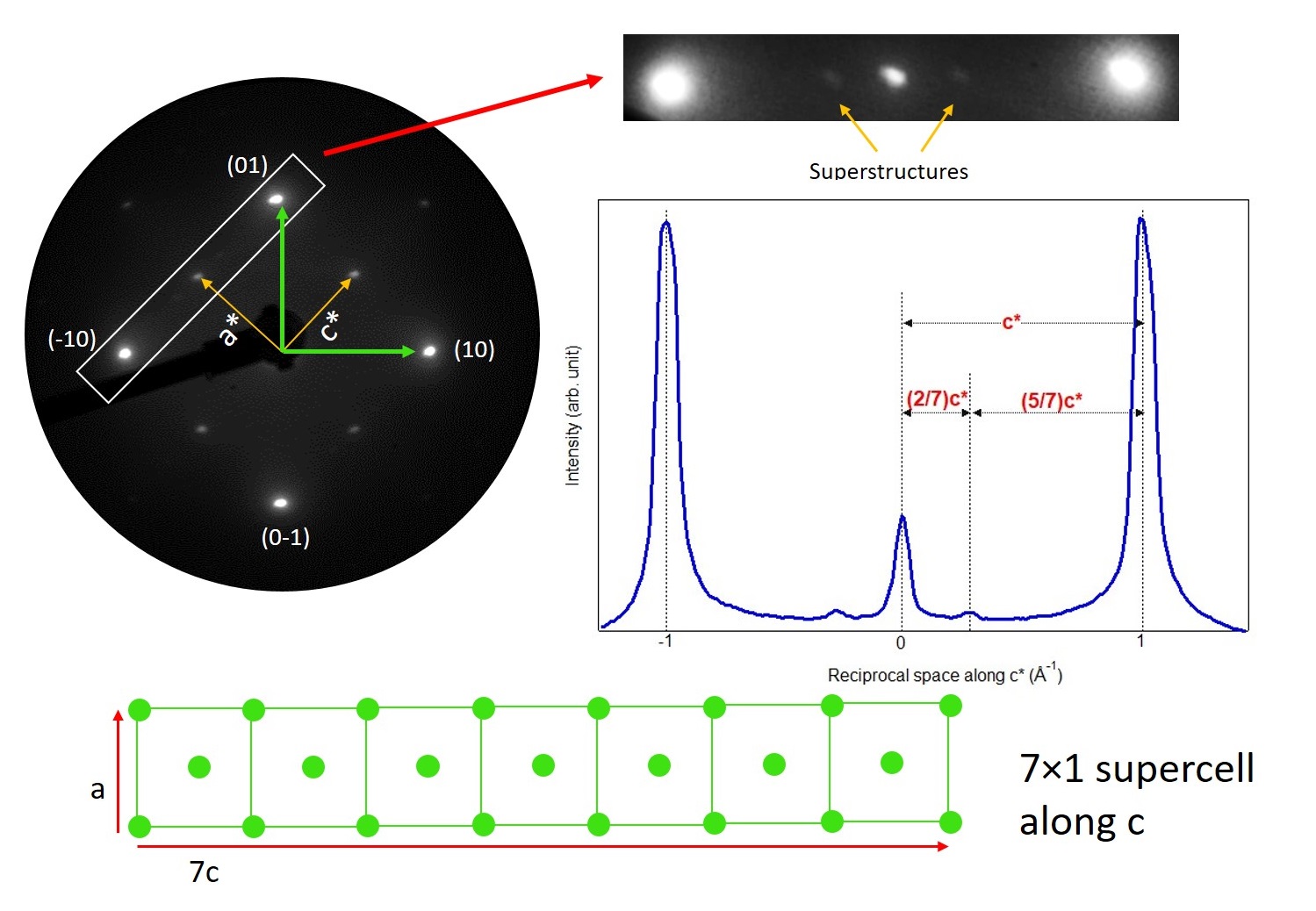}
\caption*{\textbf{Figure S3 $\mid$ Low-energy electron diffraction image of LaTe$_{3}$ single crystal.} The left panel shows low-energy electron diffraction pattern. The top-right figure is the zoomed in image of the diffraction spots. Two faint spots on both side of the central spot represent superlattice structure. The superlattice periodicity in reciprocal space is $Q = 2/7c*$, which is elaborated in bottom-left panel. This superlattice periodicity is associated to the high-temperature cdw state in LaTe$_{3}$ with the CDW transition temperature ($T_{cdw}$) presumed to be higher than 400 K. The figure at the bottom is the schematic of supercell structure along \textbf{c}-axis.}
\end{figure*}

\section{Previously known CDW transition, reflected in resistance vs temperature measurement}
Transport measurement has been done on LaTe$_{3}$ single crystal from 380 K to 430 K, and shown in Fig. S4. The temperature dependence of resistance shows a broad hump which is the clear signature of previously known CDW transition in LaTe$_{3}$
\begin{figure*}[h]
\includegraphics[width=0.8\textwidth]{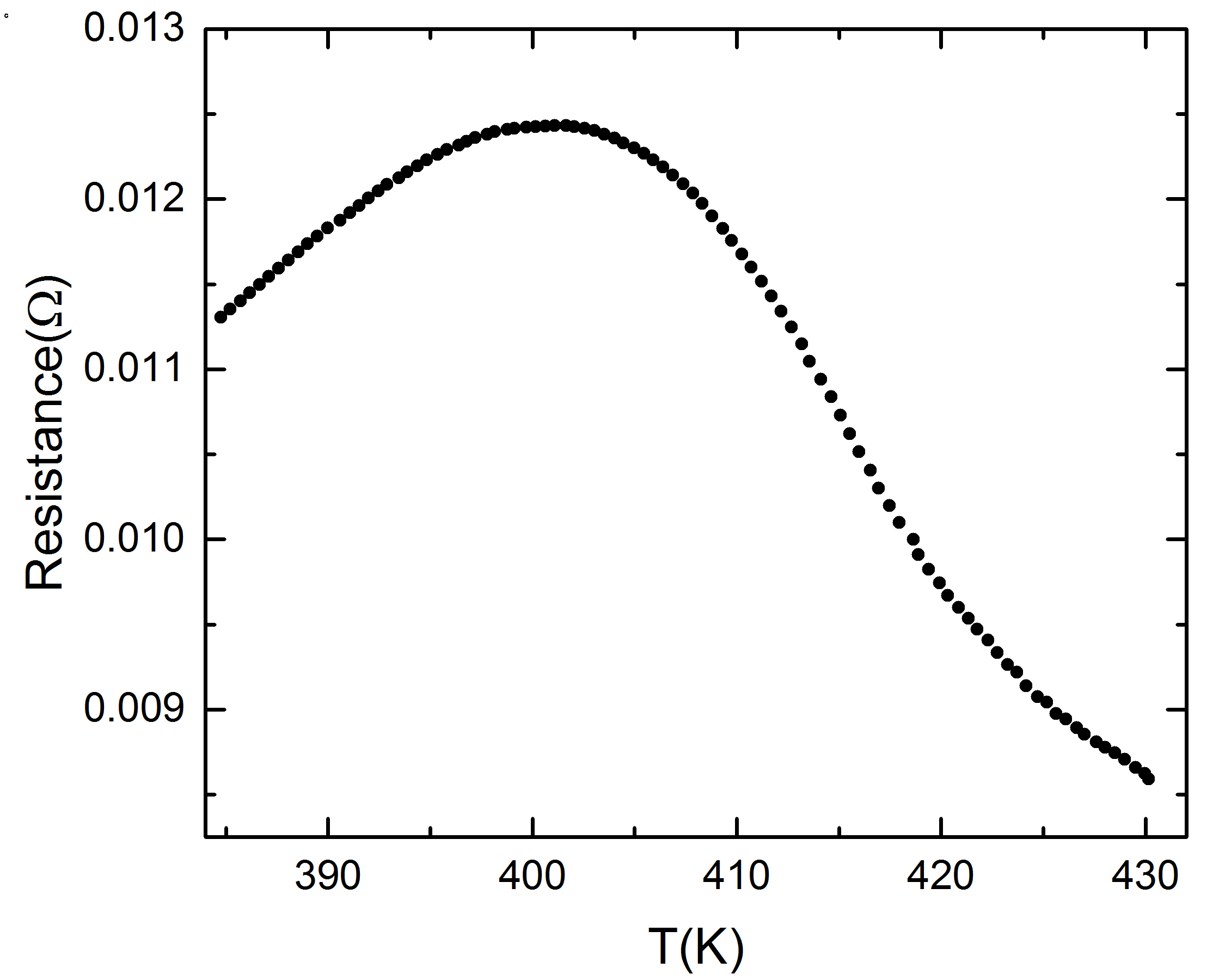}
\caption*{\textbf{Figure S4 $\mid$ Resistance as a function temperature, measured at high-temperature region.}}
\end{figure*}

\section{Field dependence of $\frac{d(MR)}{dB}$ and anisotropy ratio of TMR}
The $\frac{d(MR)}{dB}$ is linear in $B$ at low field, indicated by green solid lines (Fig. S5a). $\frac{d(MR)}{dB}$ saturates at high field, which is apparent from the data. This clearly implies that MR is quadratic in $B$ at low field and linear in $B$ at high field. With increasing temperature, the low field quadratic region in MR ($B$) increases, and as a consequence, the crossover from quadratic to linear MR shifts to higher field. In Figure S5b, we have shown the field variation of $\frac{TMR(0^{0})}{TMR(90^{0})}$ at 2 and 20 K.
\begin{figure*}[h]
\includegraphics[width=0.8\textwidth]{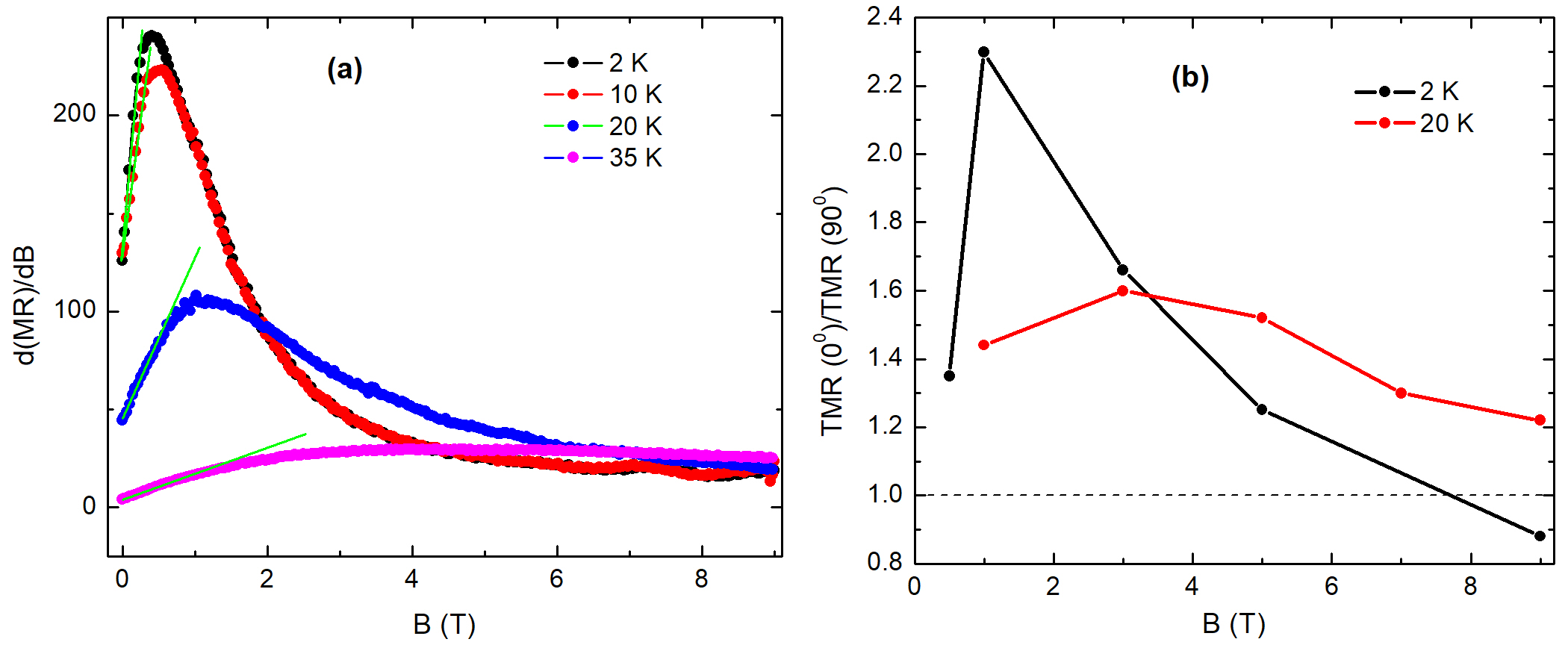}
\caption*{\textbf{Figure S5 $\mid$ $\frac{d(MR)}{dB}$ vs $B$ plot and the field dependence of MR anisotropy ratio in transverse configuration.} (a) First order derivative of MR as a function of $B$ at representative temperatures 2, 10, 20 and 35 K. The green straight line is the guide to eye, showing linear in $B$ region of $\frac{d(MR)}{dB}$. (b) The field dependence of $TMR(0^{0})/TMR(90^{0})$ at 2 and 20 K.}\label{rh}
\end{figure*}

\section{Polar plot for TMR($\theta$)}
The angle dependence of transverse magnetoresistance, which has been discussed in main text (Fig. 2), is also represented in a polar plot and shown in Fig. S6.
\begin{figure*}[h]
\includegraphics[width=0.8\textwidth]{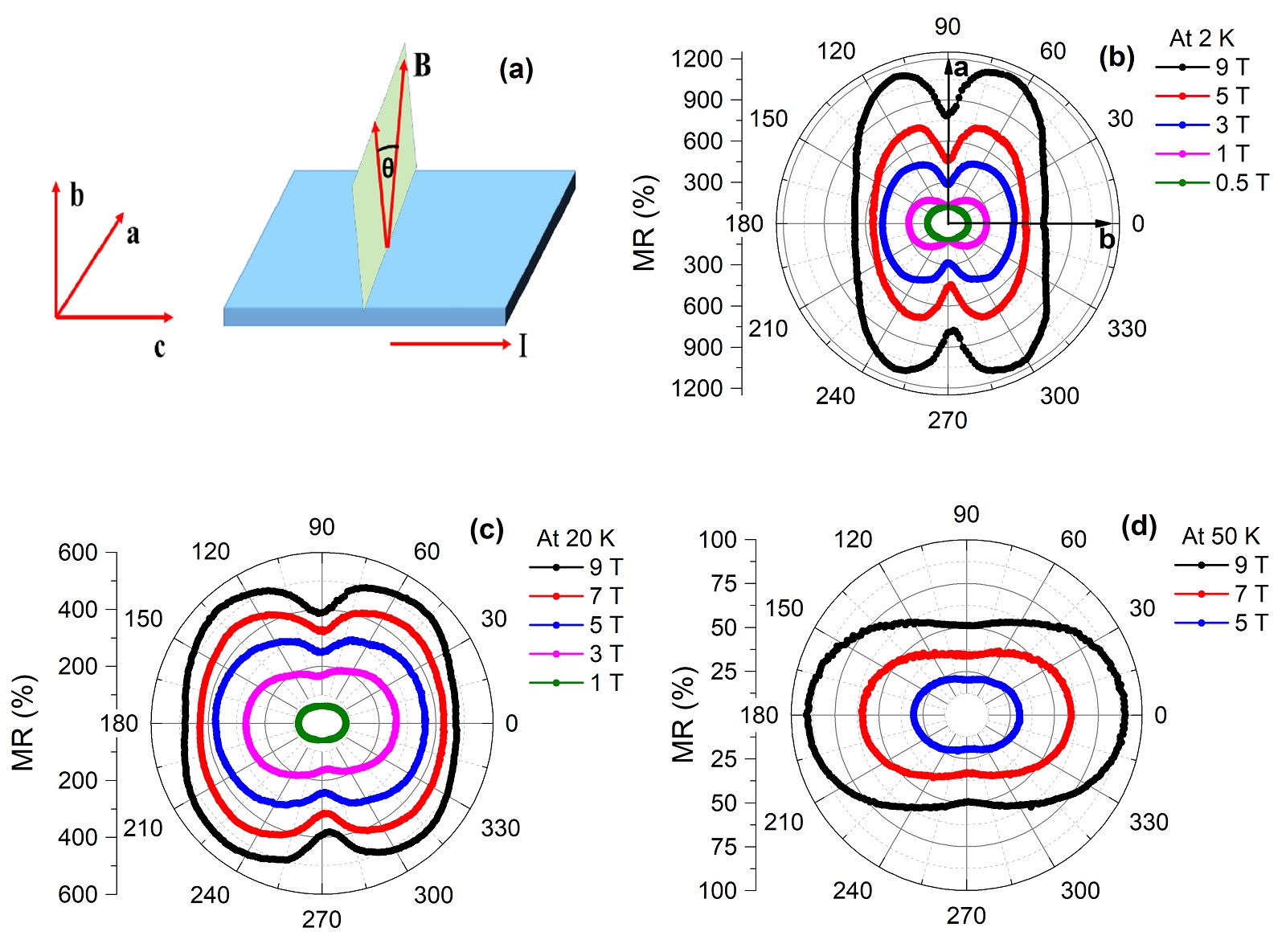}
\caption*{\textbf{Figure S6 $\mid$ Polar plot of TMR ($\theta$).}(a) The experimental configuration for the angular variation of TMR. (b), (c) and (d) are the polar plots of TMR ($\theta$) at 2 K, 20 K and 50 K, respectively. }\label{rh}
\end{figure*}

\section{Theoretically simulated Fermi surface under 1 T magnetic field}
Modification of Fermi surface under external magnetic field has been discussed in main text. Fig. 7(b-c) show the theoretically simulated Fermi surface at 3 T. In Fig. S7, we have shown the Fermi surface at 1 T both in-plane and out-of-plane configuration of magnetic field.
\begin{figure*}[h]
\includegraphics[width=0.8\textwidth]{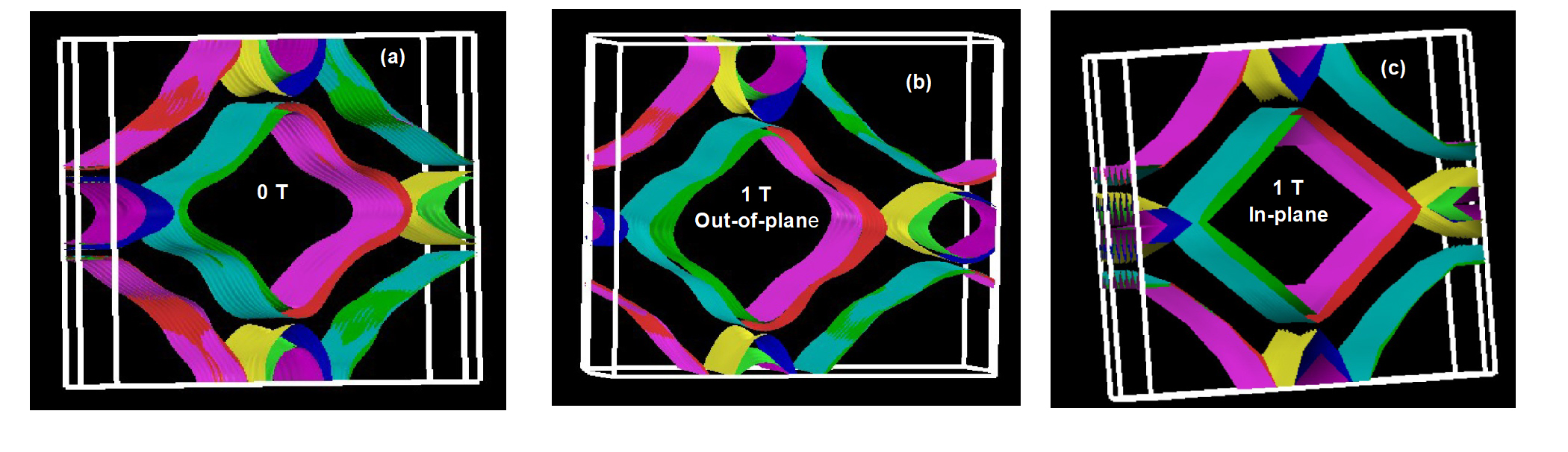}
\caption*{\textbf{Figure S7 $\mid$ Reconstruction of Fermi surface under 1 T out-of-plane and 1 T in-plane magnetic field.} Fermi surface (a) at zero field, (b) at 1 T out-of-plane, and (c) at 1 T in-plane external magnetic field.}\label{rh}
\end{figure*}

{99}

\end{document}